\documentclass[prl,twocolumn,10pt,aps, superscriptaddress]{revtex4-1}

 \usepackage{bm}
 \usepackage{verbatim}
 \usepackage{amsmath}
 \usepackage{amssymb}
 \usepackage{amsthm}
 \usepackage{latexsym}
 \usepackage{amsfonts}
 \usepackage{epsfig}
 \usepackage{epstopdf}
 \usepackage{wasysym}
 \usepackage{subfigure}
 \usepackage{color}
 \definecolor{darkblue}{rgb}{0,0,.5}
 \usepackage[linktocpage, colorlinks=true, linkcolor=darkblue, citecolor=darkblue]{hyperref}
 \usepackage[all]{hypcap}
 \usepackage[makeroom]{cancel}
 
 \usepackage[dvipsnames]{xcolor}

 \newcommand{\beginsupplement}{%
        \setcounter{table}{0}
        \renewcommand{\thetable}{S\arabic{table}}%
        \setcounter{figure}{0}
        \renewcommand{\thefigure}{S\arabic{figure}}%
        \setcounter{equation}{0}
        \renewcommand{\theequation}{S\arabic{equation}}%
     }

\begin{document}

\title{Nanoelectromechanical rotary current rectifier}

\author{Christopher W. W\"achtler}
\email{cwaechtler@pks.mpg.de}
\affiliation{Max Planck Institute for the Physics of Complex Systems, N\"othnitzer Strasse 38, 01187 Dresden, Germany}
\author{Alan Celestino}
\affiliation{Max Planck Institute for the Physics of Complex Systems, N\"othnitzer Strasse 38, 01187 Dresden, Germany}
\author{Alexander Croy}
\affiliation{Institute for Materials Science and Max Bergmann Center of Biomaterials, TU Dresden, 01069 Dresden, Germany}
\author{Alexander Eisfeld}
\affiliation{Max Planck Institute for the Physics of Complex Systems, N\"othnitzer Strasse 38, 01187 Dresden, Germany}

\begin{abstract}

Nanoelectromechanical systems (NEMS) are devices integrating electrical and mechanical functionality on the nanoscale.
Because of individual electron tunneling, such systems can show rich self-induced, highly non-linear dynamics. 
We show theoretically that rotor shuttles, fundamental NEMS without intrinsic frequencies, are able to rectify an oscillatory bias voltage over a wide range of external parameters in a highly controlled manner, even if subject to the stochastic nature of electron tunneling and thermal noise. Supplemented by a simple analytic model, we identify different operational modes of charge rectification. Intriguingly, the direction of the current depends sensitively on the external parameters.

\end{abstract}

\maketitle

\emph{Introduction.---}Rapidly growing technological abilities have resulted in fabricating devices with sizes below micrometers \cite{RugarGruetterPRL1991, JoachimEtAlNature2000, HeEtAlNanoLett2008, SubramanianEtAlNano2009, NaikEtAlNatNano2009, OroszlanyEtAlNano2010, BartschEtAlNano2012}. 
The ability to control charge transport is crucial for the design of such devices. One basic functionality is current rectification, i.e., the regulation of the preferential direction of the  current resulting from an oscillatory bias voltage. This phenomenon is based on a nonlinear current voltage characteristics. Paradigmatic candidates are NEMS, which combine mechanical and electronic degrees of freedom in a single nanoscale device \cite{CraigheadScience2000, EkinciRoukesRSI2005, BustosEtAllPRL2013, AlcazarEtAlRB2015}. So-called electron shuttles are particular examples of NEMS, where a movable island transfers electrons between two leads to which it is tunnel-coupled \cite{GorelikEtAlPRL1998, NovotnyEtAlPRL2003, NovotnyEtAlPRL2004, DonariniEtAlNJP2005, FlindtEtAlPhysicaE2005, KimEtAllPRL2010, PradePlateroPRB2012, KimEtAlPRL2013}. Notably, the rate of tunneling depends on the position of the shuttle which allows for self-induced dynamics. 
Hence, electron shuttles are  interesting candidates to investigate rectification as they do not require any active regulation from the outside.

Different realizations of electron shuttles include a harmonic oscillator \cite{GorelikEtAlPRL1998, shekhter2007nanomechanical, ScheibleBlickNJP2010} or a  rotor \cite{WangVukoivPRL2008,SmirnovEtAlNanotech2009,CroyEisfeldEPL2012, CelestinoEtAlNJP2016,WaechtlerPRAppl2019} to which the island is attached. 
While the dynamics of the former are mainly determined by the intrinsic frequency of the oscillator, the lack of such a frequency in the latter leads to complex and rich dynamics, which allow the rotor to act as a motor, sensor or electron pump \cite{CelestinoEtAlNJP2016,WaechtlerPRAppl2019}. Such a rotor could be realized by molecules on a surface \cite{EcheverriaEtAlNanoscale2014,LinEtAlJPC2019, KudernacEtAlNature2011,NattererEtAlPRL2013,JungSchlittlerGimzewskiNature1997, NickelEtAlNano2013,TierneyEtAlNatureNano2011,MishraEtAlNanoLetters2015,StipeRezaeiHoScience1998, PawinEtAl2013,EisenhutEtAlSurfaceScience2018} or an electronic island mounted onto a rigid rotor. 

While for the oscillator-based shuttle it has been shown that it is possible to rectify an oscillating signal \cite{PistolesiFazioPRL2005, ScheibleBlickNJP2010} it is not obvious whether rectification can still occur without such an intrinsic oscillatory frequency, especially in the presence of thermal noise.

\begin{figure}
 \centering\includegraphics[width=\columnwidth,clip=true]{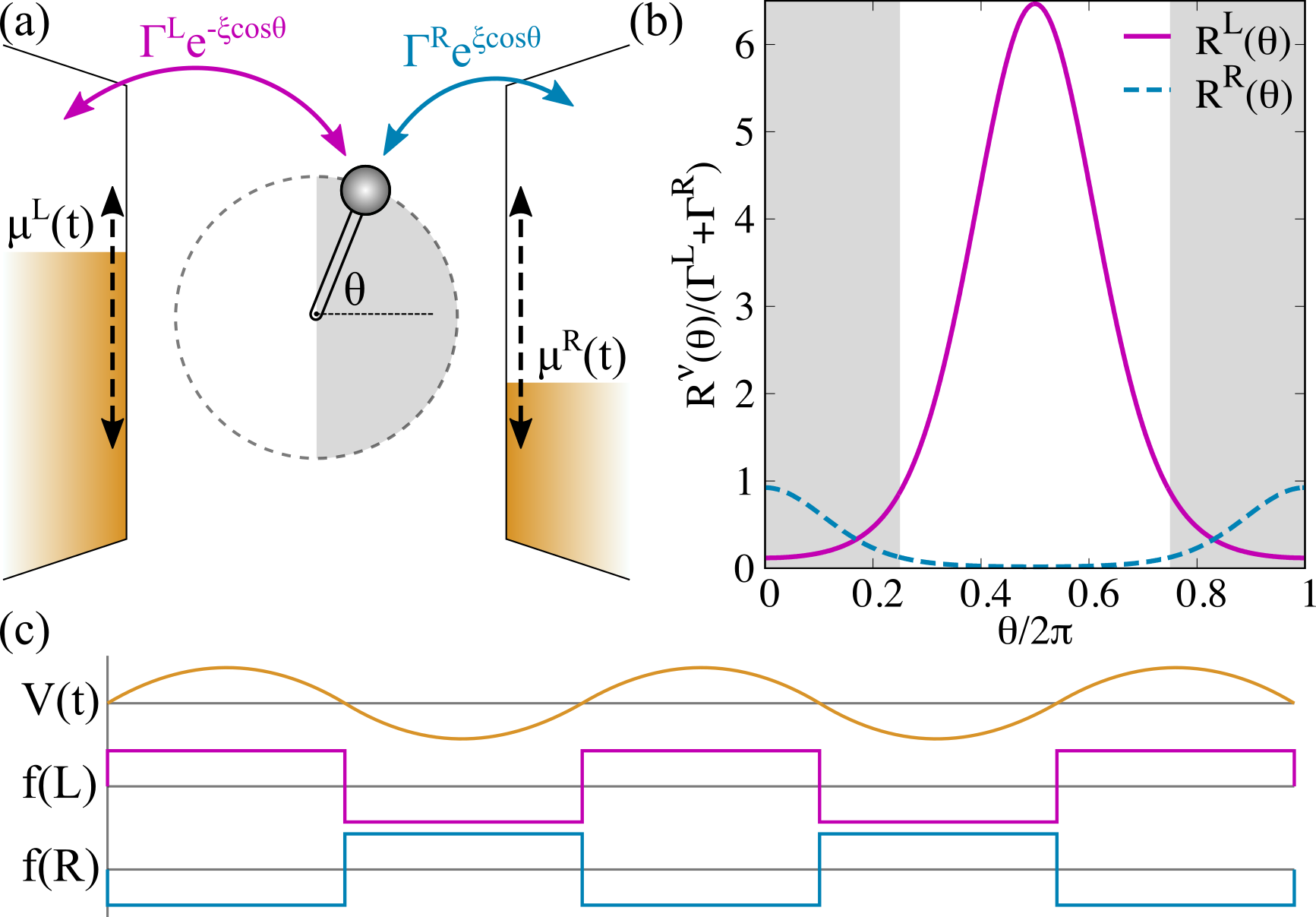}
 \label{fig:Fig1} 
 \caption{(a) Pictorial drawing of a rotor shuttle driven by time-dependent chemical potentials $\mu^\text{L/R}(t)$. (b) Tunneling rates $R^\text{L/R}(\theta)$ between system and left (magenta solid) and right (blue dashed) lead as a function of $\theta$. (c) Schematic time dependence of the oscillating voltage $V(t)$ (top) and the corresponding left (middle) and right (bottom) Fermi functions. Parameters: $\Delta = 0.75$, $\xi = 2.0$.}
\end{figure}

Here we show that rectification is also possible in the rotor, despite the absence of an intrinsic eigenfrequency. 
The rectification works over a wide range of external voltage bias magnitudes  and  frequencies. 
The rotor dynamics show a complex pattern of various operational modes  which very sensitively depends on the external parameters like friction and driving amplitude; for the same external frequency, even different directions of the current are possible. From our extensive numerical simulations, which take into account both the stochastic nature of the tunneling as well as thermal fluctuations, a clear picture of this dependence emerges which is supplemented by a simple analytic model.

\emph{Model.---}The single electron rotor, shown in Fig.~\ref{fig:Fig1}(a), consists of a single electron transistor where the central island (e.g. a quantum dot) is mounted onto a rotor \cite{CelestinoEtAlNJP2016,WaechtlerPRAppl2019}. 
The position of the island/rotor is specified by the angle $\theta$. The right-most position is taken as  $\theta = 0$  (cf.\ Fig.~\ref{fig:Fig1}(a)).
The island is subject to an electrostatic torque 
\begin{equation}
\mathcal{T}_\text{el}(\theta,q,t) = - q E(t)\sin\left(\theta\right).
\end{equation}
Here $q$ denotes the charge of the island, which we express in units of the electron charge.
The time-dependent electric field $E(t)$ is caused by a time-dependent bias voltage $E(t)=\alpha V_0\sin(\omega_\text{d} t)$, with frequency $\omega_\text d$, amplitude $V_0$ and $\alpha$ a constant factor \footnote{Approximately, $\alpha \propto 1/d$ with $d$ the distance between the electrodes}. We  assume the limit of strong Coulomb blockade \cite{GiaeverZellerPRL1968, KulikShekhter1975, AverinLikharevJLTP1986}, such that the island is either empty ($q=0$) or occupied by exactly one (excess) electron ($q=1$). This approximation is justified in the case of low temperatures, $\beta\gg 1/E_\text C$ ($E_\text C$ being the charging energy of the island), and voltages, $|V_0|\ll E_\text C$. 

Electrons can tunnel between the island and the left ($\nu=\text{L}$) and right ($\nu=\text{R}$) lead with rates $R_{q\to q'}^\nu(\theta)$, which depend exponentially on the position of the island and on the Fermi function of the respective lead. Specifically, we model the rates as \cite{SI}
\begin{equation}
\label{eq:Rates}
\begin{aligned}
R_{0\to 1}^\nu(\theta, t) &=R^\nu(\theta) f^\nu\left[-E(t)\cos(\theta)\right],\\
R_{1\to 0}^\nu(\theta, t) &= R^\nu(\theta)\left\{1-f^\nu\left[-E(t)\cos(\theta)\right]\right\}.
\end{aligned}
\end{equation}
Here  $R^{\text{L/R}}(\theta) = \Gamma^{\text{L/R}} \exp\left[\pm \xi \cos(\theta)\right]$, with dimensionless tunneling length $\xi$ \cite{CroyEisfeldEPL2012}, contains the exponential dependence of the tunnel rates. The Fermi function $f^\nu(\varepsilon) = \{1+\exp[\beta(\varepsilon-\mu^\nu)]\}^{-1}$ contains the inverse temperature $\beta$ and the charging energy of the island ($-E(t)\cos\theta$), and essentially determines the direction of tunneling via the time-dependent chemical potentials $\mu^\text{L/R}(t)=\pm V_0\sin(\omega_\text{d} t)/2$ \footnote{The system is biased symmetrically around the bare island energy such that the Fermi functions do not depend on it.}. 
For our choice of parameters, $f^{\text{L}}\approx 1$ during the first half of the driving period ($t\in(0,T/2)$) and $f^{\text{L}} \approx 0$ during the second half ($t\in(T/2,T)$), and the opposite for the right lead. 
Thus, electron tunneling is unidirectional with opposite directions during the first and second half of the driving period, respectively.

We model the effects of thermal noise on the rotor motion via Langevin dynamics. For individual trajectories the charge $q$ changes stochastically by means of a Poisson process determined by the tunneling rates (\ref{eq:Rates}) \cite{SI}. For a rotor with moment of inertia $I$ subject to friction with friction constant $\gamma$ and the electrostatic torque $\mathcal{T}_\text{el}(\theta,q,t)$, the stochastic equations of motion take the form \cite{CelestinoPHD2017, WaechtlerPRAppl2019}
\begin{equation}
\label{eq:Langevin}
\begin{aligned}
d\theta &= \omega dt, \\
Id\omega &= \left[-\gamma \omega + \mathcal{T}_\text{el}(\theta,q,t)\right] dt + \sqrt{2\gamma/\beta}dB(t),
\end{aligned}
\end{equation}
where the Wiener increment $dB(t)$ has zero mean ($\left<dB(t)\right> = 0$) and variance $\left<dB(t)^2\right>=dt$. 

To quantify rectification we use the (averaged) electron current, which is defined by the difference in the number of electrons moving to and from the right lead. A positive current corresponds to electrons tunneling from the island into the right lead ($dq^\text{R}=1$) and a negative current to the opposite direction ($dq^\text{R}=-1$). \footnote{However, the definition of the current with respect to the left lead is equivalent up to a sign because of particle conservation.} We thus calculate the current averaged over the time interval $t_\text{f} - t_\text i$ as
\begin{equation}
\left<I\right> \equiv \left<I^\text{R}\right> = \frac{1}{t_\text{f} - t_\text i}\sum dq^\text{R}.
\end{equation}
We choose the starting time $t_\text i$ and the interval $t_\text f- t_\text i$ large enough such that transient dynamics can be ignored and the current has sufficiently converged \cite{SI}.

\begin{figure*}
\includegraphics[width=0.9\textwidth]{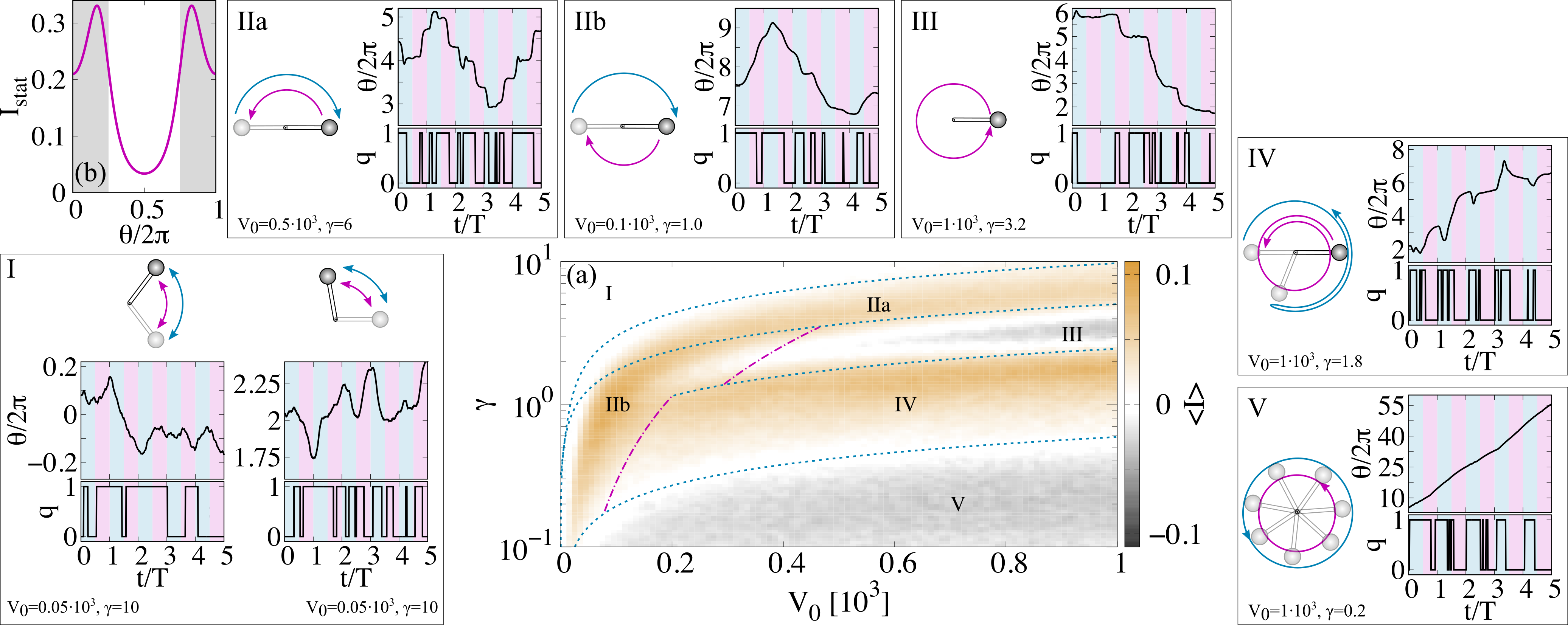}
\caption{Central panel (a): Averaged current as a function of the driving strength $V_0$ and the friction $\gamma$, where a positive value (gold) indicates rectification from left to right and a negative value (silver) from right to left. The surrounding panels show exemplary excerpts of stochastic trajectories within regimes I - V together with schematic representations of the rotor dynamics during the first (blue arrow) and second (magenta arrow) half-cycle of the driving. The dashed lines mark the borders of the individual regimes according to our analytic model. (b) Static current, when the rotor is held at a fixed angle $\theta$. Parameters: $\Delta = 0.75$, $\omega_\text{d}=1.6$, $\xi = 2.0$, $\beta = 5.0$, $\alpha = 0.1$ and $I=1.0$.}
\label{fig:Fig2}
\end{figure*}

\emph{Rectification.---}Breaking the left-right symmetry of the rotor shuttle is crucial for rectification to occur. 
We achieve such an asymmetry by choosing different  tunneling rates for the left and the right lead. The asymmetry is characterized by
\begin{equation}
\Delta = \frac{\Gamma^\text L -\Gamma^\text R}{\Gamma^\text L +\Gamma^\text R}
\end{equation}
This asymmetry-parameter has a range $-1\leq\Delta\leq1$, and in the fully symmetric case $\Delta=0$. In the following we choose $\Delta = 0.75$ to obtain a pronounced asymmetry in the tunneling rates. As clearly visible in Fig.~\ref{fig:Fig1}(b) for our parameters one has the following situation:
When the island is close to the left lead, it is essentially {\it decoupled} from the right lead and thus current through the system in either direction is significantly decreased. In the opposite case, when the island is close to the right lead (gray area) it is still coupled to the left lead (with a similar strength as to the closeby right lead) such that a current via electron tunneling is possible. Thus, in a static situation where the island is held at a fixed position rectification cannot occur.

The presence of multiple time scales - the driving freuqency $\omega_\text d$, the time scale of electron tunneling characterized by $R^\text{L/R}(\theta)$ and finally the time scale of rotation, which depends on $\gamma$ as well as on $V_0$ - promotes the rich dynamics of the rotor shuttle. 
For now, we focus on the regime where the driving frequency is comparable to the smaller electron tunneling rate, which mainly determines the electronic dynamics. We make remarks on the cases of slow and fast driving at the end of this letter. In the following, we specify $\omega_\text d / R^\text{R}(0) = 0.86$ with $\xi=2$ and analyze the impact of friction and driving strength on the system dynamics and rectification. 

In the central panel of Fig.~\ref{fig:Fig2} we show the averaged current $\left< I\right>$ as a function of the bias voltage $V_0$ and the friction $\gamma$. In the  gold shaded areas the net current is positive, i.e., on average from left to right.
In the silver shaded it is negative.
We see that rectification can be obtained over a wide range of parameters.
To understand the complex dependence on the parameters, it is instructive to analyze the underlying dynamics of the rotor.
We have identified several regions with distinct dynamic features.
These regions are marked in the figure with I--V and representative trajectories are provided in the surrounding panels.

In region I there are two  different dynamics, which both produce no net current: for small $V_0$ the island is (most of the time) located close to the right lead. As mentioned above, in this situation rectification is not possible as the island remains coupled to both leads at all times. 
For larger $V_0$ the rotor is still mostly in the vicinity of the right lead. However, during the second half-cycle (when the electric fields points from right to left) the island may move towards the left lead and pass the up(down)-right position, where one electron has been shuttled.
As the electric field changes direction, the island returns to the right lead and also shuttles one electron (from left to right). Thus, also in this situation rectification does not occur.

By decreasing the friction and/or increasing the driving strength, one enters region II. 
Here, the rotor motion is characterized by switching from left to right and vice versa during the first and second half of the driving period, respectively. In IIa these half rotations are fast compared to the driving frequency such that the island remains very close to one lead for about a quarter of the driving period. In IIb (smaller friction/smaller driving strength) the rotor turns slowly and approximately performs half a rotation during one half-cycle of the driving. 
In both cases, due to the asymmetry in the tunneling rates, electron tunneling is significantly decreased in the second half-cycle resulting in an overall positive current.

In region III the rotor remains close to the right lead during the first half of the driving period and completes one (or even two) full revolutions during the second half. Thus, in this regime the system alternates between a static and a moving mode synchronized to the driving frequency. During the full revolutions typically one electron is shuttled from right to left producing a current of $I_\text{rev}=-\omega_\text{d}/\pi$, which is larger than the current via electron tunneling in the static phase (cf. Fig.~\ref{fig:Fig2}(b)).
 Hence, there is a negative net current in this regime.

Region IV shows an intriguing interplay between friction and driving strength: During the first half-cycle the island typically moves towards the right lead, passes the closest position, loses its electron, is slowed down by friction, picks up another electron from the left lead, changes direction and approaches the right lead again. 
During this process \emph{two} electrons have been shuttled from left to right.
As the electrostatic field switches, the rotor performs one and a half rotations before arriving in the proximity of the left lead, which blocks further electron tunneling. This results in a positive average current.

Lastly, in region V the friction $\gamma$ is small, resulting in continuous rotational motion with high angular velocity. \footnote{The system rotates for long times in one direction, but may abruptly change directions and turn the opposite way for many cycles.} 
If the rotor would rotate with the same angular velocity during the first and second half-cycle, the island would be equally located close to the right as to the left lead without rectification. 
However, we find that the island is mostly occupied by an electron during the first half-cycle, such that the rotor is accelerated while turning towards the right lead. Due to the high velocity and the small tunneling rate, the occupied rotor passes the right-most position and is decelerated by the electrostatic torque and friction. 
Hence, the island during the first half-cycle  spends effectively more time close to the left lead. On the other hand, during the second half-cycle an electron on the island quickly tunnels out such that the rotor is constantly slowed down by friction, which allows for another electron to be shuttled from right to left. The uneven rotations result in an overall negative current.

\emph{Analytic model.---}From the above discussion it is apparent that the capability and direction of rectification is intimately connected to the rotational dynamics of the system. Here, the possibility of the rotor to switch between left and right positions is of great importance due to asymmetry in the tunneling rates. While electrons can tunnel through the system along the bias when the island is closer to the right lead, current is significantly decreased when closer to the left lead. Furthermore, the interplay of  multiple time scales and the presence of thermal noise contribute to the rich and complex rectification mechanism. In the following we develop a basic model of the rotor shuttle, which includes the most important features of the rectifier and promotes a clear understanding of the rectification mechanism. 

The average current mainly depends on the rotor dynamics during the second half of the driving oscillation. Consequently, of particular importance is the maximum angle  $\theta_\text{max}$ that the island can reach. From now on we will discuss the rotor dynamics during the second half-cycle when the island is initially located at $\theta(0) = 0$. As an electron tunnels into the island from the right ($q=1$) the electrostatic torque accelerates the rotor until, at  $\theta=\pi/2$, the electron tunnels out of the island into the left lead. At this moment, the rotor has an angular velocity of $\omega_\text{jump}$, such that during the remaining time $t_\text{rem}$ of the second half-cycle the empty island ($q=0$) continues to rotate while being decelerated due to friction until it stops at an final angle $\theta_\text{fin}$. We approximate $t_\text{rem}=T(\theta_\text{max}-\pi/2)/2\theta_\text{max}$. Additionally, the rotor performs a rotation of $\theta_\text{diff}=\sqrt{2t_\text{rem}/\gamma \beta}$ due to diffusion during this time span. The maximum angle the rotor can reach is then given by $\theta_\text{max}=\theta_\text{fin} + \theta_\text{diff}$, which is the central equation to approximate the transitions between the different rectification regimes. 

We now need to determine $\theta_\text{fin}$. 
To this end, we turn to Newton's equation of a deterministic damped rotor, $I \ddot \theta = -\gamma \dot \theta$ with initial velocity $\omega_\text{jump}$ and initial position $\pi/2$. 
The final angle is then given by $\theta_\text{fin} = \pi/2+\omega_\text{jump}/\gamma$. 
To estimate $\omega_\text{jump}$ we set the kinetic energy $E_\text{kin} = I\omega_\text{jump}^2 /2$ equal to the difference in potential energy of a $\pi/2$-rotation of the occupied island, which we approximate as $\Delta E_\text{el} = \alpha V_0[1-\cos(\pi/2)]$. 

Within this model, the transitions between the different dynamic regimes are now solely determined by $\theta_\text{max}$, which are for I$\leftrightarrow$II $\theta_\text{max}=5\pi/6$, for II$\leftrightarrow$III $\theta_\text{max}=3\pi/2$, for III$\leftrightarrow$IV $\theta_\text{max}=5\pi/2$ and for IV$\leftrightarrow$V $\theta_\text{max}=17\pi/2$.  We mark these transitions as blue dotted lines in the central panel of Fig.~\ref{fig:Fig2}.

However, in regime III thermal noise may prevent the rotor from approaching the right lead: When the electrostatic torque is large enough to turn the island one full rotation -- such that it would end up at the right-most position -- but the thermal noise is able push the rotor $\pi/2$ forward or backward, the island ends up on the left side again (regime IIb). 
In this case, the thermal energy $E_\text{th}= \sqrt{2\gamma/\beta}2\pi = \left<E_\text{el}\right>/4$, where $\left<E_\text{el}\right> =2 \alpha V_0 /\pi$. 
Similarly, the boundary of IV and IIb is given $E_\text{th}= \sqrt{2\gamma/\beta}3\pi = \left<E_\text{el}\right>/2$. These transitions are marked as magenta dashed lines in Fig.~\ref{fig:Fig2}.

\begin{figure}
 \centering\includegraphics[width=\columnwidth,clip=true]{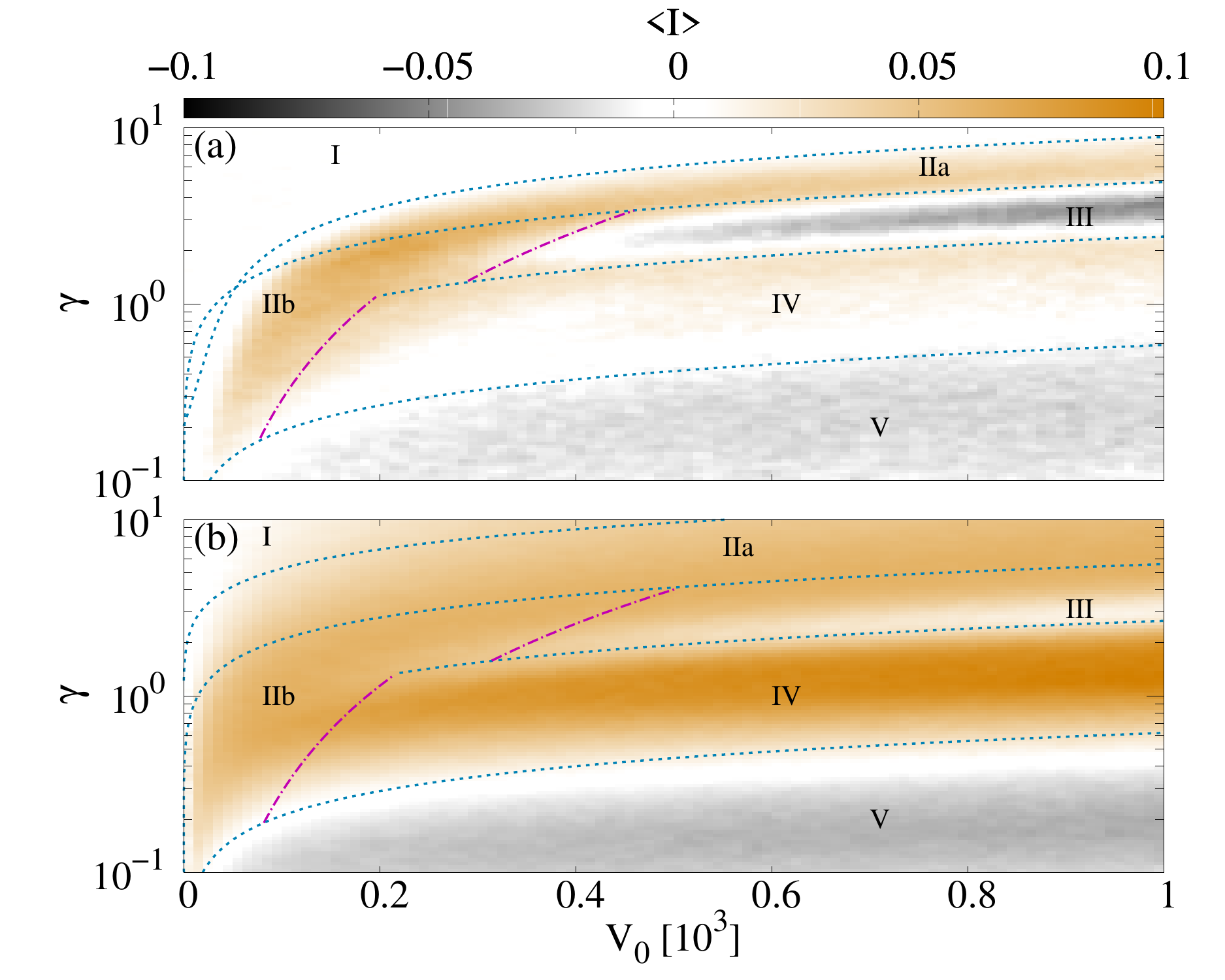}
 \label{fig:Fig3} 
 \caption{Averaged current for (a) fast driving ($\omega_\text{d}=2.5$) and (b) slow driving ($\omega_\text{d}=0.4$) compared to Fig.~\ref{fig:Fig2}. The dashed lines mark the borders of the individual regimes according to our analytic model. Parameters: $\Delta = 0.75$, $\xi = 2.0$, $\beta = 5.0$, $\alpha = 0.1$ and $I=1.0$.}
\end{figure}

\emph{Driving Frequency.---}
Increasing the driving frequency $\omega_\text{d}$ has two important consequences: as long as the rotor can follow the driving frequency, in principle more electrons can be transferred per unit time via shuttling. However, electrons have now less time to tunnel, and the rotor has less time to reach its critical angle before the field switches again. As shown in Fig.~\ref{fig:Fig3}(a), this mostly affects regimes III and IV where rectification stems from the delicate interplay of shuttling and tunneling. In regime III rectification increases as fewer electrons tunnel during the first half-cycle, while in regime IV rectification decreases as consecutive transport of two electrons during the first half-cycle becomes less likely.  
Upon further increase of the driving frequency, the rotor cannot follow the fast changes of driving direction anymore making rectification  no longer possible.

In contrast, for slow driving (cf. Fig.~\ref{fig:Fig3}(b)) more electron tunneling events can occur during each half-cycle, which enhances rectification in regime IIa and III. Similarly, for larger $V_0$ in regime I, the switching of the rotor from one side to the other induces rectification for slow driving. However, in regime III the slow driving decreases $I_\text{rev}=-\omega_\text{d}/\pi$  such that the net current becomes positive.

For both cases -- fast and slow driving -- our analytic model approximates the different rectification scenarios very well, which reaffirms the intimate relation between rectification and rotor dynamics.

\emph{Conclusions.---} We have demonstrated at the example of current rectification that a system without harmonic confinement can perform similar objectives as oscillator-based NEMS. This paves the way to study collective effects in more complex architectures, in particular, synchronization in weakly coupled systems without intrinsic frequency. 

\acknowledgements{\emph{Acknowledgements.---}We thank M. T. Eiles for fruitful discussions and for a careful reading of the manuscript. CWW acknowledges support from the Max-Planck Gesellschaft via the MPI-PKS Next Step fellowship.
AE acknowledges support from the DFG via a Heisenberg fellowship (Grant No EI 872/5-1).}

%

\clearpage
\onecolumngrid
\appendix
\beginsupplement
\section{SUPPLEMENTAL MATERIAL}

In this supplemental material we list details on the model and the numerical implementation of the stochastic differential equations (Sec.~I), details on an analytic model for the transition within regime I (Sec.~II), an investigation on the dependency of rectification on the driving frequency (Sec.~III), and an investigation on the dependency of rectification on the temperature (Sec.~IV). 

\section{I. Details on the model and the numerical implementation}
\label{sec:A}

We model the stochastic tunneling of electrons between the island and the leads via a Poisson process, i.e., the change in the occupation of the island $q$ changes according to
\begin{equation}
\label{eq:Poisson}
\begin{aligned}
dq &= \sum\limits_\nu dq^\nu =\sum\limits_{\nu q'}(q'-q)dN^\nu_{q'\to q}(\theta,t). 
\end{aligned}
\end{equation}
Here, the Poisson increments $dN^\nu_{q'\to q}(\theta, t)\in\{0,1\}$  obey the statistics $\left<dN^\nu_{q'\to q}(\theta,t)\right> = R_{q'\to q}^\nu(\theta,t)dt$ and $dN^\nu_{q'\to q}(\theta,t)dN^{\tilde \nu}_{q'\to \tilde q}(\theta,t) = \delta_{\nu\tilde \nu}\delta_{q'\tilde q}dN^\nu_{q'\to q}(\theta,t)$, which specify that the average number of jumps into state $q$ from a state $q'$ in a time interval $dt$ is given by the tunneling rate $R^\nu_{q'\to q}(\theta, t)$ and that only one tunneling event per time interval can occur, i.e., either all $dN^\nu_{q'\to q}(\theta,t)=0$ or $dN^\nu_{q'\to q}(\theta,t)=1$ for precisely one set of indices $q$, $q'$ and $\nu$. 

The tunneling rates $R^\nu_{q\to q'}(\theta,t)$ depend exponentially on the position of the island and on the Fermi function $f\left[\varepsilon(\theta, t)-\mu^\nu(t)\right]$ of the lead $\nu$. In general, the energy of the island is given by $\varepsilon(\theta, t) = \varepsilon -E(t)\cos(\theta)$, where $\varepsilon$ denotes the bare energy of the island. However, we choose a symmetrical bias with chemical potentials $\mu^{\text L/\text R}(t) = \varepsilon \pm V_0\sin(\omega_d t)/2$, such that the Fermi functions are independent of $\varepsilon$. This leads to tunneling rates of the form defined in Eq.~(2) of the main manuscript. 

For our numerical studies, we choose the initial angle, angular velocity and dot occupation from uniform distributions of the intervals $[-\pi,\pi)$, $[-2,2]$ and $[0,1]$, respectively. Our simulations run in total for 5000 driving cycles and we average the current over 4000 driving cycles, i.e., $t_f = 5000~T$ and $t_i = 4000~T$. We have checked that the dynamics is independent of the initial configuration and that the averaged current has sufficiently converged.

\section{II. Analytic model for regime I}
\label{sec:B}
In regime I there are two different dynamics, which both produce no net current: First, for small $V_0$ the rotor is (most of the time) located to close to the right lead which we call from now on regime Ia, whereas for larger $V_0$ the rotor may switch from right to left during the second half-cycle of the driving period (regime Ib). The transition from Ia to Ib is hence determined by the possibility of the rotor to arrive at a maximum angle of $\theta_\text{max} =\theta_\text{fin} +\theta_\text{diff}= \pi/2$ during the time of $T/2$. Here, $\theta_\text{diff}= \sqrt{T/\gamma\beta}$. 

Similar to the analytic model of the main manuscript, we assume that the island is located at $\theta(0)=0$ with zero initial velocity at the beginning of the second-half cycle and that is occupied by an electron ($q=1$). Furthermore, we assume that during the full half-cycle of $T/2$ the island remains occupied. To approximate $\theta_\text{fin}$ in this situation, we turn to a deterministic damped rotor which is constantly pushed by an effective electrostatic torque, i.e., $I\ddot \theta = -\gamma \dot\theta + \mathcal T_\text{eff}$. We approximate the effective electrostatic torque by the average electrostatic torque exerted on an occupied island located between $0\leq \theta\leq \pi/2$, which is given by 
\begin{equation}
\mathcal T_\text{eff} = \frac{2}{\pi}\frac{2}{T}\int\limits_0^{\pi/2}\int\limits_{T/2}^{T} \mathcal T_\text{el}(\theta, q=1,t) dt~d\theta.
\end{equation}
We will discuss this additional transition together with the transitions discussed in the main manuscript when we analyze the dependency of rectification on the driving frequency and temperature in Sec.~III and IV. 

\section{III. Dependency of rectification on the driving frequency}
\label{sec:C}

\begin{figure*}[h]
\includegraphics[width=\textwidth]{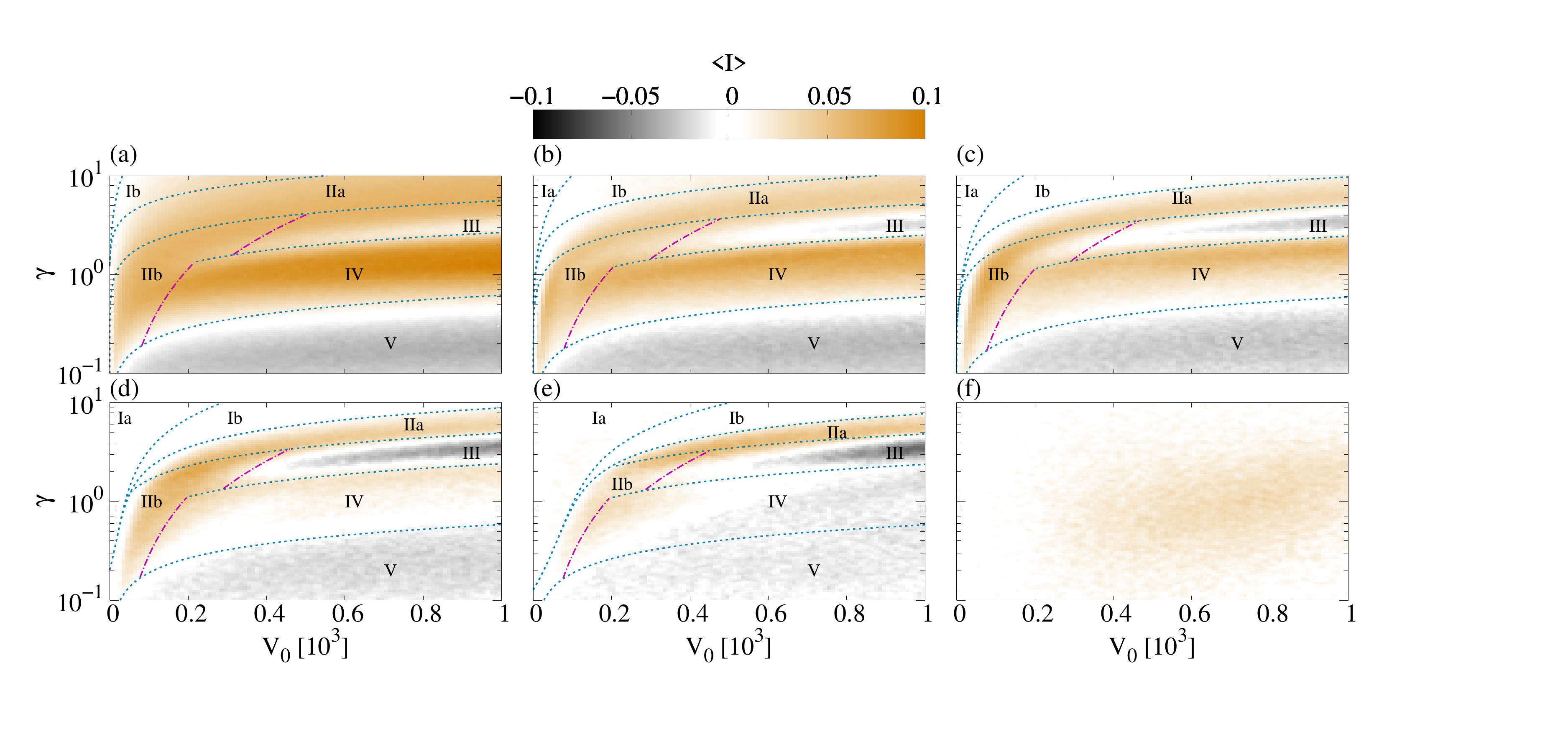}
\caption{Averaged current as a function of the driving strength $V_0$ and the friction $\gamma$, where a positive value (gold) indicates rectification from left to right and a negative value (silver) from right to left for different driving frequencies: (a) $\omega_\text{d}=0.4$, (b) $\omega_\text{d}=1.0$, (c) $\omega_\text{d}=1.6$, (d) $\omega_\text{d}=2.5$, (e) $\omega_\text{d}=4.0$, and (f) $\omega_\text{d}=15.0$. The dashed lines mark the borders of the individual regimes according to our analytic model(s). Parameters: $\Delta = 0.75$, $\xi = 2.0$, $\beta = 5.0$, $\alpha = 0.1$ and $I=1.0$.}
\label{fig:FigSI_1}
\end{figure*}

In this section, we discuss the effect of changing the driving frequencies on the different rectification regimes. In Fig.~\ref{fig:FigSI_1} we show the averaged current $\left<I\right>$ as function of the driving strength $V_0$ and the friction $\gamma$ for different values of $\omega_\text{d}$. Here, panel (c) corresponds to the driving frequency of Fig.~\ref{fig:Fig2} of the main manuscript ($\omega_\text{d} = 1.6$), whereas panels (a) and (b) show the current for slower driving and panels (d) -- (f) for faster driving. 

As discussed in the main manuscript, for slower driving (Fig.~\ref{fig:FigSI_1}(a) and (b)) more electron tunneling events may take place during each half-cycle. As a result, regime III becomes smaller and vanishes for very small driving frequencies. Since in the first half-cycle the rotor is located close to the right lead, many electrons can tunnel during the increased driving period. In the second half period all electrons are shuttled. Due to the increased driving period, current via tunneling in the static phase is larger than shuttling a single (or even multiple) electron(s). 

On the other hand, in regime IV rectification becomes more pronounced for slow driving frequencies. Similar as to regime II, in the first half-cycle, where the rotor is located close to the right lead, many electrons can tunnel, while in the second half-cycle the rotor ends up on the left, which prevents further tunneling resulting in a larger positive net current. 

In regime Ib we observe rectification upon decreasing the driving frequency. As discussed in Sec.~II in this regime the rotor switches from right to left during the second half-cycle, where electron tunneling is significantly decreased. As the driving period becomes large, the time the rotor spends close to the left lead becomes longer such that the asymmetry in the tunneling rates affects the net current. Thus, this regime exhibits rectification for small driving. 

We now discuss the case of faster driving relative to panel (c): Upon increasing the driving frequency the golden regions IIa and IV become narrower (in gamma direction) and the silver region III becomes more pronounced and appears at smaller driving strength. In contrast to the previously discussed slow driving case, current via electron tunneling becomes less likely. Hence, in regime III very few electron tunnel during the first half-cycle and at least one electron is shuttled in the second half-cycle which increases rectification in this regime. 

In regime IV, as the driving period decreases, the available time to pick up another electron from the left and release it to the right also decreases resulting in a smaller net current (panel (d)). Upon further increase of the driving frequency the net current in regime IV even becomes negative for larger $V_0$. Here, one electron is picked up from the left during the first half-cycle but not released to the right during the short time the rotor is located close to the right lead such that the same electron tunnels into the left lead again during the second-half cycle. However, it may happen that the empty rotor picks up another electron during the second half cycle, which then again gets shuttled back and forth. However, this results in a small negative net current.

For a very large driving frequency (panel (f)) rectification is significantly decreased as the rotor dynamics is unable to follow the fast switching of the electrostatic field. Nevertheless, we see a small positive net current for large values of $V_0$ suggesting that the position of the island is not totally independent of the driving field direction. In fact, we observe a small asymmetry in the distribution of  rotor positions during the first and second half-cycle -- closer to the left lead when the electric field is pointing from right to left -- which due to the asymmetric tunneling rates results in a small positive net current. However, for frequencies larger than $\omega_d\gtrsim 20.0$ rectification essentially vanishes. Overall, we observe that rectification is possible over a large range of driving frequencies and especially also occurs for small driving frequencies. 

\section{IV. Dependency of rectification on the temperature}
\label{sec:D}

\begin{figure*}[h]
\includegraphics[width=\textwidth]{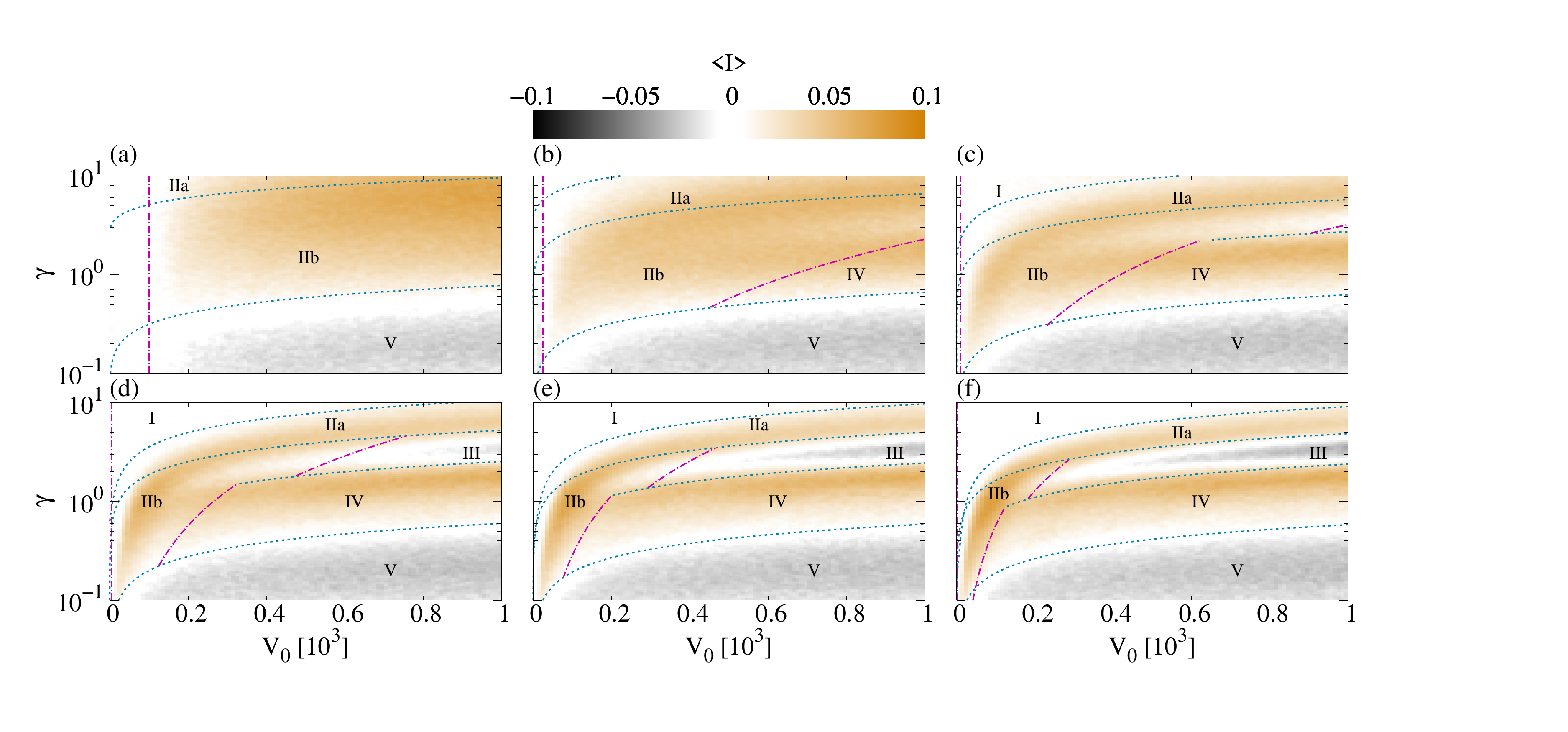}
\caption{Averaged current as a function of the driving strength $V_0$ and the friction $\gamma$, where a positive value (gold) indicates rectification from left to right and a negative value (silver) from right to left for different temperatures: (a) $\beta=0.1$, (b) $\beta=0.4$, (c) $\beta=1.0$, (d) $\beta=2.5$, (e) $\beta=5.0$, and (f) $\beta=10.0$. The dashed lines mark the borders of the individual regimes according to our analytic model. Parameters: $\Delta = 0.75$, $\omega_\text{d}=1.6$, $\xi = 2.0$, $\alpha = 0.1$ and $I=1.0$.}
\label{fig:FigSI_2}
\end{figure*}

In this section we discuss the effect of changing the (inverse) temperature $\beta$ on the rectification observed in Fig.~\ref{fig:Fig2} of the main manuscript, which is the same as panel (e) in Fig.~\ref{fig:FigSI_2}. We observe that by changing the temperature the transitions between the different regimes are shifted. The direction of these shifts can be understood from our basic analytic model: The temperature affects the free diffusion angle $\theta_\text{diff} = \sqrt{2t_\text{rem}/\gamma\beta}$. Hence, the blue dotted lines are shifted downwards by increasing $\beta$. Additionally, as the temperature is decreased ($\beta$ increased) the area where the rotor dynamics is significantly affected by the thermal noise (regime IIb) decreases in $V_0$ direction. Moreover, for large temperatures there exists a regime where the rotor dynamics is essentially equal to that in equilibrium as the equilibrium thermal energy is larger than the electrostatic energy. Here, rectification does not occur. This regime is located in the small $V_0$ regime and upper bounded by $\alpha V_0 = 1/2\beta$ which we mark by the vertical magenta dashed line in Fig.~\ref{fig:FigSI_2}. Overall, we observe that rectification is possible in a large range of temperatures. 

\end{document}